\documentstyle[12pt]{article}

\begin{document}
\renewcommand{\thesection}{\arabic{section}.}
\renewcommand{\theequation}{\thesection\arabic{equation}}

\titlepage

\def\lng{{\rm ln}\; {\cal G}}
\def\lin{{\rm ln}}
\def\lug{{\rm ln \; g}}
\def\lu{{\rm ln }}
\def\Kahler{K\"{a}hler}
\def\eeq{\end{equation}}
\def\beq{\begin{equation}}
\def\Tr{{\rm Tr}}
\def\k{\zeta}
\def\G{{\cal G}}

\def\beqa{\begin{eqnarray}}
\def\eeqa{\end{eqnarray}}
\def\kahler{K\"{a}hler}
\def\lag{Lagrangian}
\def\Lag{Lagrangian}
\def\D{ {\cal D}}
\def\L{ {\cal L}}
\def\h{ {\cal H}}
\def\R{ {\cal R}}
\def\C{ {\cal C}}
\def\Z{ {\cal Z}}
\def\x{\times}
\def\ra{\rightarrow}

\begin{flushright} UPR-0502T \end{flushright}
\vspace{4ex}

\begin{center} \bf

COORDINATE AND K\"{A}HLER $\sigma$-MODEL ANOMALIES AND THEIR
CANCELLATION IN STRING EFFECTIVE FIELD THEORIES \\
      \rm
\vspace{3ex}
Gabriel Lopes Cardoso \\ and \\
Burt A. Ovrut\footnote{Work supported in part by the
Department of Energy under Contract No. DOE-AC02-76-ERO-3071
and NATO Grant No. 860684.} \\
Department of Physics \\ University of Pennsylvania \\ Philadelphia, PA
19104-6396 \\

\vspace{3ex}

ABSTRACT

\end{center}

We discuss the complete set of one-loop triangle graphs involving
the Yang-Mills gauge connection, the \Kahler\ connection and the
$\sigma$-model coordinate connection in the effective field
theory of $(2,2)$ symmetric $Z_N$ orbifolds.  That is, we discuss
pure gauge, pure \Kahler\ and pure $\sigma$-model coordinate
anomalies as well as the mixed anomalies, such as \Kahler-gauge,
some of which have been discussed elsewhere.  We propose a
mechanism for restoring both \Kahler\ and $\sigma$-model
coordinate symmetry based upon the introduction of two types of
counterterms.  Finally, we enlarge the $\sigma$-model generalization
of the Green-Schwarz mechanism to allow the removal of
the universal parts of a wider class of anomalies than those
previously discussed.

\newpage

\section{Introduction}

\hspace*{.3in} \Kahler\ superspace provides a powerful and elegant
tool for the
description of the $N =1$ supergravity-matter system~[1] in
four dimensions.  The structure group of \Kahler\ superspace
contains an abelian group, denoted $U_K(1)$, in addition to the
Lorentz group.  By construction, $N=1$ supergravity-matter
systems are invariant under \Kahler\ transformations.  These
consist, in part, of field dependent $U_K (1)$ transformations of
chiral component field fermions.  In addition to the \Kahler\
symmetry, the tree-level supergravity-matter theory
possesses
several other invariances.  A generic matter manifold with
\Kahler\ metric $g_{i\bar{j}}$ is parameterized by chiral matter
fields $\phi^i$ and $\bar{\phi}^{\bar{j}}$~[2].  Under $\sigma$-
model general coordinate transformations of these fields, the
tree-level Lagrangian and, hence, the action transform like
scalars.  It follows that the partition function is, naively,
invariant under $\sigma$-model coordinate transformations.
Under certain circumstances, the isometries of the \kahler\
metric $g_{i \bar{j}}$ may be actual symmetries
of the tree-level component Lagrangian.
Another invariance of relevance
for this paper is the invariance under Yang-Mills
transformations.  Hence, the covariant derivative $\D_m$ of a
generic matter Weyl spinor $\chi^i_\alpha$ contains connections
$a_m$, $\Gamma^i\,_{jk}$ and $v^{(r)}_m $ for gauging
\Kahler\ transformations, $\sigma$-model coordinate
transformations and Yang-Mills gauge transformations,
respectively.  We will not discuss space-time Lorentz
transformations               and, hence, we will not consider
the Lorentz connection $\omega_m\,^\alpha\,_\beta$ in the main
part of this paper.  Mixed $\sigma$-model
anomalies involving this connection
have been
considered in a recent paper~[3], and we will briefly summarize
and expand these results in Appendix~A.
Given these three symmetries at tree-level,
one might ask whether or not they are anomalous at the one-loop
level.  As always in four dimensions, one-loop fermionic
triangle graphs
are potentially anomalous.  In such a triangle graph massless
fermions run around the loop and any of the three gauge
connections under consideration, $a$, $\Gamma$, and $v$, can
couple to each of three vertices of the triangle graph.  Hence,
the symmetries under consideration can be broken by pure $aaa$,
$\Gamma\Gamma\Gamma$ and $vvv$ graphs as well as by any of the
mixed graphs, such as $avv$ or $\Gamma vv$.  In this paper, we
will focus on $(2,2)$ symmetric $Z_N$ orbifolds~[4,5].  The
effective theory of these
orbifolds is, by construction~[4], free of pure
Yang-Mills gauge
anomalies, pure Lorentz anomalies and mixed Yang-Mills, Lorentz
anomalies.  Therefore, we need not consider
the pure $vvv$ graph in our discussion.
One has, however, to worry about anomalies in
\Kahler\ and $\sigma$-model coordinate transformations.
Anomalies in $\sigma$-model coordinate transformations, as
computed from $\Gamma \Gamma \Gamma$ graphs, have been studied
in other contexts in~[6].  Recently, much attention has been
devoted to the study of the two
mixed graphs $avv$ and $\Gamma vv$, in the context of non-harmonic
gauge coupling constants~[3,7,8,9,10]. It was
shown~[3,8,9] that these two mixed graphs contribute
non-local terms to the 1-loop effective \lag, which are not
invariant under either \Kahler\ transformations
or under modular
transformations (a subset of isometry transformations).  Clearly,
it is not enough to only consider the two mixed $avv$ and $\Gamma vv$
graphs when discussing the issue of anomalies under \Kahler\
and $\sigma$-model coordinate transformations.  In this
paper, we will compute all anomalous contributions, from both the
pure $aaa$ and $\Gamma \Gamma \Gamma$ graphs as well as from all
the mixed graphs, such as $avv$ and $\Gamma vv$. We will find
that all these graphs contribute non-local terms to the 1-loop
effective \lag, which either break \Kahler\ invariance or
transform as densities under $\sigma$-model coordinate
transformations, and, hence, are anomalous under these
transformations.
That is, the
1-loop effective \lag, as computed from the massless fields in
the $Z_N$ orbifold theory, breaks \Kahler\ invariance and is also
not a scalar under $\sigma$-model coordinate  transformations.
This state of affairs is devastating to making physical sense of
orbifold theories.  It follows that one would like to devise some
cure to render such theories anomaly free.  It has been shown in
explicit string amplitude calculations~[7] that, in the case
of mixed $avv$ and $\Gamma vv$ graphs, the integration over heavy
orbifold states contributes local counterterms to the 1-loop
effective \lag\ of the light fields, so as to restore modular
invariance of the gauge coupling constants~[9].  These
explicit string calculations then suggest a general
mechanism, based upon
the introduction of heavy state induced
counterterms, for rendering the theory
anomaly free.  Hence, we will follow a similar strategy and
postulate the existence of two types of counterterms for the
restoration of invariance of the partition function
under \kahler\ transformations and
under $\sigma$-model coordinate transformations.
The first type of counterterm
will be added to the anomalous effective 1-loop \lag\ so as to
restore invariance under \kahler\ transformations.  This
counterterm is of the type of a superpotential term for the
untwisted moduli and the dilaton.  A second type of counterterms
will be added to the anomalous \lag\ so as to
render the effective Lagrangian a scalar under
$\sigma$-model coordinate transformations.  In order to
fix the explicit form of all these counterterms we will choose the
standard coordinate system for parametrizing the manifold of the
untwisted moduli and the dilaton.
In the standard coordinate system, superstring amplitudes are
expected to be modular invariant~[11].
Hence, the effective theory for orbifolds
should be modular invariant as well.  By demanding modular
invariance we will be able to uniquely determine the moduli
dependence of
the superpotential term as well as of the type two
counterterms.  Additional computations of three
particle scattering string amplitudes, such as three dilaton
amplitudes, would be required in order to determine the
dilaton dependence of the
counterterms.  Finally, it has been
shown~[10] that the universal
part of the mixed $avv$ and
$\Gamma vv$ anomalies gets removed by a $\sigma$-model
generalization~[8] of the Green-Schwarz mechanism~[12]
involving the dilaton, dilatino, axion linear supermultiplet.
We will close this paper by discussing an extension of this
mechanism.  Specifically we will show that the universal
part of the pure \Kahler\ anomaly,
$aaa$, can, in principle, be removed as well by a Green-Schwarz type
mechanism.  However, only an explicit calculation of the relevant
three point correlation function in string theory can clarify
whether or not a part of the pure \Kahler\ anomaly
(and how much of it) is
removed by this mechanism in string theory.

\section{$\bf U_K (1)$ Superspace}

\hspace*{.3in} \Kahler\ superspace provides a powerful and elegant
tool for the description of the supergravity-matter system~[1].  We
begin by recalling here some of the features of \Kahler\ superspace
geometry which will be relevant in the subsequent discussion.  A
complete description of the properties of $U_K(1)$ superspace can
be found in~[1].  The structure group of \Kahler\
superspace is taken to be $SL(2,\C) \x U_K (1)$ and, accordingly,
one introduces two Lie algebra valued one-form gauge connections
$\phi_B\,^A = dz^M \phi_{MB}\,^A$ and $A= dz^M$ $A_M$
corresponding to the Lorentz and $U_K(1)$ groups respectively.
In addition, one introduces a supervielbein $E_M\,^A$ and the
associated one-forms $E^A = dz^M E_M\,^A$.  The $U_K(1)$ gauge
connection $A$ is a composite gauge connection defined by
\beqa              A_\alpha &=& \frac{1}{4} \D_\alpha K  \nonumber \\
A^{\dot{\alpha}} &=& - \frac{1}{4} \bar{\D}^{\dot{\alpha}} K  \nonumber    \\
A_{\alpha \dot{\alpha}} &=& - \frac{i}{8} \left[ \D_\alpha,
\bar{\D}_{\dot{\alpha}} \right] K
\eeqa 
where the prepotential $K(\Phi_i,\; \Phi^{+}_i)$ is the \Kahler\
potential for matter fields.  All matter superfields have
vanishing $U_K(1)$ weight, $\omega(\Phi_i) = 0$.  Under a
\Kahler\ transformation
\beq K ( \Phi_i, \; \Phi_i^+) \ra K(\Phi_i, \; \Phi^+_i) +
F(\Phi_i) + \bar{F} (\Phi_i^+) \eeq 
the one-form $A$ transforms as
\beq A \ra A + \frac{i}{2} d \: Im F \eeq 
Also, under a \Kahler\ transformation the supervielbein one-forms
$E^A$ can be shown~[1] to transform as
\beq E^A \ra E^A \exp \left[ - \frac{i}{2} \omega (E^A) Im F
\right] \eeq 
where
\beq  \omega (E^\alpha) = 1, \: \: \omega (E_{\dot{\alpha}}) = -1,
\: \: \omega(E^a) = 0        \eeq 

Solving the Bianchi identities subject to a set of natural
constraints~[13], one finds that all components of the
torsion and curvature may be expressed in terms of a set of
superfields and their coordinate derivatives:
\beq \begin{array}{cccccc} {\rm superfield} & R & R^+ & G_{\alpha
\dot{\alpha}} & W_{\alpha \beta \gamma}, X_\alpha &
\bar{W}_{\dot{\alpha} \dot{\beta} \dot{\gamma}}, \bar{X}_{\dot{\alpha}}
          \\
U_K(1) \; {\rm weight} & 2 & -2 & 0 & 1 & -1 \end{array}
\eeq 
where
\beqa                     X_{\alpha} &=& \D_{\alpha} \R -
\bar{\D}^{\dot{\alpha}} G_{\alpha \dot{\alpha}} = - \frac{\kappa^2}{8}
\left( \bar{\D}^2 - 8R\right) \D_{\alpha} K \nonumber \\
\bar{X}^{\dot{\alpha}} &=& \bar{\D}^{\dot{\alpha}} R^+
+ \D_{\alpha} G^{\alpha \dot{\alpha}}
= - \frac{\kappa^2}{8} \left( \D^2 - 8R^+ \right)
\bar{\D}^{\dot{\alpha}} K  \eeqa 
and ${\kappa}^2 = 8\pi M^{-2}_P$ , where $M_P$ is the Planck mass.
$X_\alpha$ is the superfield fieldstrength of the $U_K(1)$ gauge
connection.

\section{Classical Symmetries of (2,2) Symmetric Orbifold
Lagrangians}

\setcounter{equation}{0}

\hspace*{.3in} In this section, we will consider (2,2) symmetric
$Z_N$ orbifolds~[4,5] with gauge group $G = E_8 \otimes E_6 \otimes
U(1)^2$.  These are the orbifolds which have been discussed most
extensively in the context of non-harmonic gauge coupling
constants in superstring theories.  We will, for simplicity,
furthermore restrict our discussion to the subset of these
orbifolds which do not contain (1,2) moduli; that is, to $Z_7, \;
Z_8$ and $Z_{12'}$.  Our considerations can, however, be
generalized in a straightforward way to the case of the remaining
$E_8 \otimes E_6 \otimes U(1)^2$ orbifolds, $Z_{6'}, \; Z_{8'}$
and $Z_{12}$, which do contain (1,2) moduli.  Generically, the
spectrum~[4] of the orbifolds under consideration contains
three uncharged untwisted (1,1) moduli, $T^I$, as well as untwisted and
twisted {\bf27} matter fields, which we will be collectively denoting
by $\phi^i$.  We also consider twisted (1,1) moduli, denoted
$C_d$.  However we will ignore matter fields which are singlets
under the $E_6$, since the structure of their \Kahler\ potentials
is unknown.  In addition to the above fields, we will also
discuss the dilaton field $S$~[14].  Each of the untwisted
(1,1) moduli $T^I$ parametrizes a coset space $SU(1,1) / U(1)
\simeq SL(2,\R)/ U(1)$ with isometry group $SL(2, \R)$~[15].
The \Kahler\ potential for the moduli and matter fields under
consideration reads~[5,16]
\beqa             K_M &=& \kappa^{-2} \sum_I - ln (T + \bar{T})^I
+ \sum_i e^{ln {\cal G}_i } \bar{\phi}^i e^V \phi^i  \nonumber \\
&& + \sum_d e^{ln {\cal G}_d} \bar{C}_d e^V C_d + {\cal O} \left( \left(
\bar{\phi} \phi \right)^2 \right) + {\cal O} \left( \left( \bar{C} C
\right)^2 \right) \eeqa 
where
\beqa           {\cal G}_i \left(T^I, \bar{T}^I\right) &=& \prod_I
\left[ T^I + \bar{T}^I \right]^{- q_i^I} \nonumber\\
               {\cal G}_d \left(T^I, \bar{T}^I\right) &=& \prod_I
\left[ T^I + \bar{T}^I \right]^{- q_d^I} \eeqa 
The exponents $q^I_i$ and $q^I_d$ are associated with the matter
fields $\phi^i$ and twisted moduli fields $C_d$,
respectively~[16].  The \Kahler\ potential for the universal
dilaton supermultiplet, present in any compactification scheme of
the heterotic superstring theory, is given by~[14]
\beq K_S = - \kappa^{-2} ln (S + \bar{S}) \eeq 
where $S$ is the chiral representation of the dilaton.  We have
chosen to normalize the fields $T^I$ and $S$ in units of Planck
mass.

The classical kinetic \lag\ for the supergravity-matter-dilaton
supermultiplet system is~[1]
\beq {\cal L}_0 = - 3 \kappa^{-2} \int d^4 \theta E \eeq 
with \Kahler\ potential $K=K_M + K_S$~[1].
The kinetic Yang-Mills \Lag\ is given by~[1]
\beq    \L_{YM} = \frac{1}{8} \int d^4 \theta \frac{E}{R} S W^\alpha
W_\alpha + h.c.  \eeq 
where $W^\alpha$ denotes the Yang-Mills superfield fieldstrength.
The gauge coupling constant $g$ is, at tree-level, given by the lowest
component of $S$, $g^{-2} = \left. Re S \right|$.  The
complete kinetic \Lag,
\beq \L = \L_0 + \L_{YM} \eeq 
as well as the potential energy part of the Lagrangian and
the partition function
possess the following classical invariances. \\  \\
1. \Kahler\ invariance: \\
This follows from the transformation law of $E$, given in (2.4),
under \Kahler\ transformation (2.2).  \\    \\
2. Gauge invariance:   \\
This follows from the fact that both $E$ and $R$ are functions of the
\Kahler\ potential $K$ which, in turn, is invariant under
Yang-Mills transformations of the charged fields $\phi^k$
and $C_d$
\beqa             e^V &\ra& e^{-i \bar{\Lambda}} e^V e^{i
\Lambda} \nonumber \\
\phi^k &\ra& e^{-i \Lambda} \phi^k \nonumber \\
C_d &\ra& e^{-i \Lambda} C_d  \eeqa 
where
\beq \bar{\D}^{\dot{\alpha}} \Lambda = 0    \eeq 
and where $V$ denotes the Yang-Mills prepotential vector
superfield.  The square of the Yang-Mills superfield fieldstrength,
$W^\alpha W_\alpha$, is by construction invariant under
Yang-Mills transformations.   \\   \\
3. $\sigma$-model coordinate invariance and
$SL(2, \R)$ isometry invariance:  \\
$\sigma$-model coordinate invariance of the partition function
follows from the fact that, under these transformations, the
Lagrangian and action, though not in general invariant,
behave as scalars.  When integrated over $\int \ldots
[d \phi^i] [d \bar{\phi}^{\bar{j}}] \sqrt{\det g_{i \bar{j}}}$
in the path integral this assures that the partition function
is invariant under $\sigma$-model coordinate transformations.
The action of the $SL(2, \R)$ isometry group, a subset of
$\sigma$-model coordinate transformations, on a modulus $T^I$
can be represented by~[15]
\beq T^I \ra \frac{aT^I - ib}{icT^I + d}\eeq 
where
\beqa      ad-bc = 1, \:\:\:    a,b,c,d \in \R\eeqa 
If the transformations (3.9) are accompanied by~[15]
\beqa
\phi^i & \ra & \left( \prod_I \left( i c T^I + d \right) ^{-q^I_i}
\right) \phi^i = e^{-\sum_I q_i^I F^I} \phi^i \nonumber \\
C_d &\ra & \left(\prod_I \left( i c T^I + d \right)^{-q^I_d}
\right) C_d = e^{-\sum_I q^I_d F^I} C_d \nonumber \\
S &\ra& S \eeqa
where
\beq F^I = ln (i c T^I + d), \eeq 
then $K$ transforms as
\beqa      \kappa^2  K \ra \kappa^2  K
+ F (T) + \bar{F} (\bar{T}) \eeqa 
with
\beqa F(T) = \sum_I F^I \eeqa 
and, hence, the superdeterminant $E$ is invariant.  It follows
that the complete tree-level kinetic Lagrangian (3.6)
is invariant under $SL(2, \R)$.  However, the potential
energy part of the classical Lagrangian generically
breaks $SL(2, \R)$ invariance.
The transformations (3.9) and (3.11) on the coordinates $(T^I,
\phi^i, C_d)$ should be regarded as special continuous coordinate
transformations on the $(T^I, \phi^i, C_d)$-manifold.  Note that
the modular transformations are the subset of these isometry
transformations with $a$, $b$, $c$, $d$  $\in \Z$.  That is,
the modular transformations correspond to the action of the
$SL(2, \Z)$ subgroup of $SL(2, \R)$.  The entire tree-level
Lagrangian is $SL(2, \Z)$ invariant, since orbifold potential
energies respect this symmetry.

Invariances (1) - (3) of the tree-level theory of the
supergravity-matter-dilaton supermultiplet system can, of course,
also be displayed at component level.  Component fields are
defined according to standard notation~[1]: $A^i, \;
\chi_{\alpha}^i, \; {\cal F}^i$ for chiral multiplets (and similar
notations for antichiral multiplets) and $\lambda_\alpha,
\;v_m, \; \D$ for Yang-Mills multiplets.  The covariant
derivative of a generic Weyl fermion will, in a theory with
invariances (1) - (3), then contain a connection for each of these
symmetries. \\    \\
1. The connection for gauging \Kahler\ transformations (2.2)
is given by the lowest component~[1] of the $U_K(1)$ gauge
connection superfield $A_{\alpha \dot{\alpha}}$ in (2.1)
\beqa             \left. A_{\alpha \dot{\alpha}} \right| &=&
a_{\alpha \dot{\alpha}} = \sigma^m_{\alpha \dot{\alpha}}
a_m \nonumber \\
a_m &=& \frac{1}{4} \left( \partial_jK \D_m A^j - \partial_{\bar{j}}
K\D_m \bar{A}^{\bar{j}} \right) +
 i\frac{1}{4} g_{i\bar{j}}\left( \chi^i \sigma^m
\bar{\chi}^{\bar{j}}\right) \eeqa 
\\  \\
2. The Yang-Mills connection $v_{\alpha \dot{\alpha}} =
v^{(r)}_{\alpha \dot{\alpha}} T^{(r)}$ (where $T^{(r)a}\,_b$
denote the generators of the Yang-Mills gauge group $G$) is
contained in the Yang-Mills prepotential $V$
\beq \frac{1}{2} \left. \left[ \D_\alpha, \bar{\D}_{\dot{\alpha}}
\right] V \right| = -2 v_{\alpha \dot{\alpha}} \eeq 
\\  \\
3. The connection $\Gamma^i\,_{jk}$ which insures covariance
under general coordinate transformations on a generic $\sigma$-model
manifold of metric $g_{i \bar{j}}$ also insures covariance under
isometries of the metric $g_{i\bar{j}}$, since these are a
special subset of coordinate transformations.  $\Gamma^i\,_{jk}$
is given by~[2]
\beq \Gamma^i\,_{jk} = g^{i \bar{j}} \partial_j g_{k\bar{j}}
\eeq 
For the \Kahler\ potential (3.1) at hand, we find that, for the
matter superfields $\phi^{i}$
\beq \Gamma^i\,_{Jk} = \delta^i\,_k \partial_J \ln g_i \eeq 
where
\beq ln {\cal G}_i | = ln g_i  \eeq
Therefore, the complete covariant derivative for matter fermions
$\chi_{\alpha}^{i a}$ reads~[1]
\beqa             \D_m \chi^i &=& \partial_m \chi^i - \kappa^2 a_m
\chi^i + (\partial_m T^J \Gamma^i\,_{Jk}) \chi^k  \nonumber \\
&+& i v_m^{(r)} (T^{(r)} - \frac{1}{2} D^{(r)}) \chi^i +
\cdots  \eeqa 
where
\beq D^{(r)} = \kappa^2 \frac{\partial K}{\partial A^{ia}}T^{(r)a}\,_b
A^{ib} \eeq 
The dots stand for additional gravitational couplings such as the
coupling to the space-time Lorentz connection
$\omega_m\,^\alpha\,_\beta$.
Such additional couplings have been considered in a recent
paper~[3], and we will not include them here in our main
discussion. They are, however, briefly commented on
in Appendix~A of this paper.
  Note that the connections $a_m$ and
$\Gamma^i\,_{jk}$ are two totally distinct geometrical objects,
as made explicit by the different powers of $\kappa^2$ multiplying
them in in (3.20).

Of relevance for this paper are the couplings of the quantum
currents to the connections discussed above, which will be
taken to be external fields.  We will consider quantum superfield
fluctuations around a classical, $SL(2, \R)$ preserving background
$T^I \neq 0, \; \phi^i = 0, \; C_d = 0, \; S \neq 0$.  This is consistent
with the fact that higher order terms in $\phi^i$ and $C_d$ have been
dropped in (3.1).  Collecting all the relevant interaction terms
in the component \lag~[1] we find the couplings of the
quantum matter current $\bar{\chi}^{\bar{j}c}_{\dot{\alpha}} \;
\chi^{ia}_\alpha$ to be
\beqa             \L_{int} &=& - \frac{i}{2}
\bar{\chi}^{\bar{j}}_{\dot{\alpha} a} \left( \delta_{i \bar{j}} g_i
\bar{\sigma}^{m{\dot{\alpha}\alpha}} X_m\,^a \,_b\,_i
\right) \chi^{ib}_\alpha +
{\cal O} \left( \left( \chi^i \bar{\chi}^{\bar{j}} \right)^2 \right)
\eeqa 
where
\beqa             X_m\,^a\,_b\,_i &=& - 2 \kappa^2 a_m \,
\delta^a\,_b  \nonumber  \\
&& + \left( \partial_m T^I \partial_I ln g_i \right. - \partial_m
\bar{T}^{\bar{I}} \partial_{\bar{I}} ln g_i
+ i \partial_I \partial_{\bar{J}} ln g_i \left.\, \chi^I
{\sigma}^m \bar{\chi}^{\bar{J}} \right) {\delta}^a \,_b  \nonumber \\
&& + 2 i v_m^{(r)} \; T^{(r) a} \, _b \eeqa 
Note that the Killing potentials $D^{(r)}$ given in (3.21) do
not appear in (3.23), since they vanish for the classical
$SL(2,\R)$ preserving background we have chosen.  It will be
useful to write $X_m\,^a\,_b\,_i$ as the $\theta
\bar{\theta}$-component of a prepotential superfield $Y$.  It
follows from (3.15), (3.16), and (3.19) that
\beq X_m\,^a\,_b\,_i = \frac{i}{4} \bar{\sigma}_m
^{\dot{\gamma}\gamma} \left. \left[ \D_\gamma,
\bar{\D}_{\dot{\gamma}} \right] Y^a\,_b\,_i \right| \eeq 
where
\beqa             Y^a\,_b\,_i &=& Z_i \delta^a\,_b + V^a\,_b \nonumber \\
Z_i &=& -\kappa^2 \frac{1}{2} K + ln \G_i \eeqa 
Similarly, one finds that the couplings of the twisted quantum
modulino currents $\bar{\chi}_{\dot{\alpha} d} \; \chi_{\alpha d}$
and the quantum gaugino
currents $\bar{\lambda}^b_{\dot{\alpha}} \lambda^a_\alpha $ to the
external gauge connections are again as in (3.24), where the
prepotential superfields are now respectively given by
\beqa
 Y^a\,_b\,_d &=& Z_d \delta^a\,_b + V^a\,_b  \nonumber\\
 Y^a\,_b\,_V &=& Z_V \delta^a\,_b + V^a\,_b \eeqa
where
\beqa
 Z_d &=& - \kappa^2 \frac{1}{2} K + ln \G_d  \nonumber\\
 Z_V &=& \kappa^2 \frac{1}{2} K \eeqa 
Note that, due to the specific Yang-Mills kinetic term (3.5),
there is an additional coupling of $Im\, S|$ to the divergence of
the gaugino current.  $Im\, S|$, however, does not transform
under \kahler\ transformations.  This is so, because superfield
$S$ in (3.5) is inert under \kahler\ transformations.  In fact,
superfield $S$ coordinatizes the dilaton manifold and, hence, the
coupling of $Im \,S|$ to the divergence of the dilatino current
in (3.5) should under no circumstances be reinterpreted as an
additional coupling of the gaugino current to any of the
connections under consideration.  That is, this coupling is of no
relevance for the discussion of $\sigma$-model anomalies and,
hence, we will ignore it.
In an analogous manner one finds that the couplings of the
untwisted quantum modulino currents $\bar{\chi}^I_{\dot{\alpha}}
\chi_\alpha^I$ to the external connections are contained in $X_{mI}$
which can be written as the $\theta \bar{\theta}$-component of a
prepotential superfield as
\beq X_{m I} = \frac{i}{4} \bar{\sigma}_m ^{\dot{\gamma} \gamma}
\left. \left[ \D_\gamma, \bar{\D}_{\dot{\gamma}} \right] Z_I
\right|  \eeq 
where
\beq Z_I = - \kappa^2 \frac{1}{2} K + ln \G_I \eeq 
and
\beq \G_I = \left. \frac{\partial^2 K_M}{\partial T \,^I
\partial {\bar{T}} \,^{\bar{I}}} \right|_{\langle \rangle}
\eeq 
where $|_{\langle \rangle}$ means evaluation in the $SL(2, \R)$
preserving background.
Finally, the couplings of the twisted quantum
dilatino current are as in (3.29), where the
prepotential superfield is now given by
\beqa
Z_S &=& - \kappa^2 \frac{1}{2} K + ln \G_S \eeqa 
where
\beq \G_S =\left. \frac{\partial^2 K_S}{\partial S
\partial {\bar{S}}}\right|_{\langle  \rangle}  \eeq 
Note that no mixed quantum currents, $\bar{\chi}_{\dot{\alpha}} ^i
\chi^I_\alpha$, between matter fermions and untwisted modulinos
occur for the background chosen, and similarly for twisted
modulinos.

\section{K\"{a}hler and $\bf \sigma$-model Coordinate Anomalies
in the Effective Theory}

\setcounter{equation}{0}

\hspace*{.3in} In this section, we begin by evaluating non-local one-loop
corrections associated with triangle graphs in the theory of the
$E_8 \otimes E_6 \otimes U(1)^2$ orbifolds discussed in Section
3.  We will work to lowest order in prepotentials $Z$ and $V$.
These non-local terms will, in general, be anomalous under the
symmetry transformations discussed in the previous section.  Some
of these terms can be straightforwardly generalized to the case
of (2,2) orbifolds with larger gauge group than $U(1)^2$, such as
$Z_3$.  As an example, the non-local term containing the \Kahler\
prepotential $K$ only, can easily be generalized to the
off-diagonal moduli case.
For concreteness, we will, in Appendix A, discuss
the pure \Kahler\ anomaly for the $Z_3$ case, which
we will find to be non-vanishing.  This is the term generated by
the triangle graph in which all three vertices couple to $K$.

We begin by considering matter fields $\phi^k$ and the associated
prepotential superfield $Y_k$ given by (3.25).  For the classical
background chosen, the external \Kahler\ prepotential $K$ in
(3.25) is given by
\beq \kappa^2 K = - \sum_I ln (T^I + \bar{T}^I) - ln (S + \bar{S})
\eeq 
The transformation law of the prepotential superfied $Y_k$ can
then be found from (2.2), (3.7) and (3.11), and reads, to
linearized order,
\beqa             \delta Y_k &=& \delta Z_k + \delta V \nonumber \\
\delta Z_k &=& i \Lambda_k - i \bar{\Lambda}_k \nonumber\\
\delta V &=& i \Lambda - i \bar{\Lambda} \eeqa 
where
\beq i \Lambda_k = - \frac{1}{2} \kappa^2 F(T,S) +
 \sum_I ( - \frac{1}{2} + q^I_k ) F^I
\eeq 
and
\beq \bar{D}^{\dot{\alpha}} \Lambda^k = \bar{D}^{\dot{\alpha}}
\Lambda =0. \eeq 
The classical \lag\ (3.6) is invariant under gauge
transformations (4.2).  It is of interest to ask whether or not
the quantum corrected effective action is invariant under (4.2)
as well.  As always in four space-time dimensions, one-loop
non-local anomalous terms in the effective \lag\ are associated
with three-point triangle graphs~[17].  The relevant
supergraph is shown in Figure~1.  In this graph, matter
superfields $\phi^k$ run around the triangle loop and
prepotential superfield $Y_k$ couples to each of the three
matter field vertices, as discussed in Section 3.  A similar
graph has been considered in~[18] in the context of
supersymmetric, non-abelian chiral anomalies.  There~[18]
it is the Yang-Mills prepotential superfield $V$ which couples to
each of the three matter field vertices.  The graph in [18]
contributes a complicated non-local term to the effective
action that has not been computed in~[18].  Its variation
under a Yang-Mills transformation, $\delta e^V$, however, has been
computed in~[18] in a field theoretical way as an infinite
power series in $(e^V -1)$ using a Pauli-Villars regularization
scheme.  This result can be easily transcribed to our case, since
the graph in Figure~1 is similar to the one discussed
in~[18], as mentioned.  One first expands the result
in~[18] to lowest order in prepotential $V$, and
subsequently replaces $V$ and its variation $\delta V = i \Lambda
- i \bar{\Lambda}$ in [18] by $Y_k$ and its variation
$\delta Y_k$ (as given by (4.2)).  If we further take into
account that the orbifold models are, by construction~[4], anomaly
free under  Yang-Mills transformations, then the relevant
transformation of prepotential superfield $Y_k$ is given by
the parameters in (4.3) only.  Thus,
we arrive at the following finite and anomalous
expression for the variation of the effective action $\Gamma$
under (4.3).  It is
\beq \delta \Gamma_{\rm matter} = \sum_k (i \Lambda_k \circ G - i
\bar{\Lambda}_k \circ \bar{G}), \eeq 
where
\beqa
i \Lambda_k \circ G &=& \frac{1}{(4 \pi)^2} \frac{1}{(48)}
\int d^4 \theta i \Lambda_k  \left[ \right.
n_k^{E_6} (\bar{D}_{\dot{\alpha}} D^\alpha Z_k) (
\bar{D}^{\dot{\alpha}} D_\alpha Z_k)  \nonumber \\
&+&  \Tr_k (3 (\bar{D}_{\dot{\alpha}}
D^\alpha V^a\,_c)(\bar{D}^{\dot{\alpha}} D_\alpha V^c\,_b)
+ 3 (\bar{D}_{\dot{\alpha}} D^\alpha V^a\,_b) (
\left. \bar{D}^{\dot{\alpha}} D_\alpha Z_k)) \right] \eeqa 
where $n^{E_6}_k = \Tr^{E_6}_k \delta^a\,_b$.  The
relative factor of 3 is due to Bose symmetrization in the $ZZZ$-channel.
We would like to rewrite expression (4.6) in terms of superfield
fieldstrengths defined as follows.
The Yang-Mills superfield fieldstrength is, to lowest order in
prepotentials, given by
\beq W_\alpha = - \frac{1}{8} \bar{D}^2 D_\alpha V \eeq 
We defined another superfield fieldstrength with $Z_k$ as its
prepotential field
\beq W_{\alpha k} = - \frac{1}{8} \bar{D}^2 D_\alpha Z_k = -
\frac{1}{2} X_\alpha - \frac{1}{8} \bar{D}^2 D_\alpha ln {\cal G}_k \eeq
where $X_{\alpha}$ is given by (2.7).  It is instructive as well
as usefull to display the component fieldstrengths contained in
$X_{\alpha}$ and ${\bar{D}}^2 D_{\alpha} ln {\cal G}_k$.  These
component fieldstrengths are contained in the $\theta$-component
of $X_{\alpha}$ and ${\bar{D}}^2 D_{\alpha} ln {\cal G}_k$,
respectively
\beqa    X_\alpha |_{\theta_{\beta}} &=& {\delta}_{\alpha}^{\beta}
\kappa^2 D^K (y) - ({\sigma}^m {\sigma}^n)_{\alpha}^{\beta}
{\kappa}^2 v^K_{mn}(y)   \nonumber \\
- \frac{1}{8} {\bar{D}}^2 D_{\alpha} ln {\cal G}_k |_{\theta_
{\beta}}
&=& {\delta}_{\alpha}^{\beta} D_k (y) -
({\sigma}^m {\sigma}^n)_{\alpha}^{\beta} v_{mn}^k (y) \eeqa
In the above expression, $v_{mn}^K$ denotes the fieldstrength of
the \Kahler\ connection $a_m$ given in (3.15), $v_{mn}^K = (
{\partial}_m a_n - {\partial}_n a_m)$, and $v_{mn}^k$ contains
the fieldstrength of the $\sigma$-model manifold connection (3.18)
pulled back to space-time
\beqa  v_{mn}^k &=& - \frac{1}{2} R_{I \bar{J}}^k (\partial_m
T^I \partial_n \bar{T}^{\bar{J}} - (m \leftrightarrow n)) \nonumber\\
&+& \frac{i}{4} {\partial_m ((\chi^I \sigma_n \bar{\chi}^{\bar{J}})
R_{I \bar{J}}^k ) - (m \leftrightarrow n)}  \eeqa 
where
\beq R_{I \bar{J}}^k = \partial_{I} \partial_{\bar{J}} ln {\cal G}_k
\eeq 
The $D$-terms in (4.9) are given ~[1] by
\beqa D^K &=& ( - g_{I \bar{J}} \partial_m T^I \partial^m
\bar{T}^{\bar J} - g_{S \bar{S}} \partial_m S \partial^m \bar{S}
+ g_{I \bar{J}} {\cal F}^I {\cal F}^{\bar{J}} + g_{S \bar{S}}
{\cal F}^S {\cal F}^{\bar{S}})   \nonumber\\
D_k &=& ( - R_{I \bar{J}}^k \partial_m T^I \partial^m
\bar{T}^{\bar{J}} + R_{I \bar{J}}^k {\cal F}^I {\cal F}^{\bar{J}} )
\eeqa  
Expression (4.6) can then be rewritten as
\beq i \Lambda_k \circ G = - \frac{1}{3} \frac{1}{(4 \pi)^2} \int
d^2 \theta i \Lambda_k \left[ n^{E_6}_k
W^\alpha_k W_{\alpha k}
+ 3 \Tr_k W^\alpha W_\alpha + 3 (\Tr_k W^\alpha) W_{\alpha k}
\right]     \eeq 
The right hand side is non-vanishing and, hence, $\delta \Gamma_
{\rm matter}$ given in (4.5) is non-zero.  Therefore, the theory is
anomalous under (4.3).  Expression (4.13) can be obtained by
variation of the following non-local term in the effective \lag
\beqa             \Gamma_{\rm matter} &=& \frac{1}{12} \frac{1}
{(4 \pi)^2} \sum_k \int d^4 \theta \left[ n^{E_6}_k W^\alpha_k
W_{\alpha k} \right.  \nonumber \\
&& + 3 \Tr_k (W^\alpha W_\alpha)\left. + 3 (\Tr_k W^\alpha)
W_{\alpha k} \right] \frac{1}{\Box} D^2 Z_k + h.c. \eeqa 
The anomalous contributions to the effective theory from the
untwisted and twisted moduli superfields and the dilaton
superfield running around a triangle loop can be calculated in a
similar way.  Each of these contributions can be readily
obtained from (4.14) in the following way.  As in (4.8), we first
define superfield fieldstrengths $W_{\alpha I}$, $W_{\alpha d}$
and $W_{\alpha S}$ of the prepotential superfields $Z_I$, $Z_d$,
and $Z_S$.  Then, the contributions from the moduli and dilaton
superfields are obtained from (4.14) by replacing $W_{\alpha k}$
and $Z_k$ with the corresponding $W$ and $Z$ for each of these
fields.  The contributions read
\beqa
\Gamma_{untwisted \,moduli}
&=& \frac{1}{12} \frac{1}{(4 \pi)^2} \sum_I \int
d^4 \theta
W^\alpha_I W_{\alpha I} \frac{1}{\Box} D^2 Z_I + h.c. \nonumber \\
\Gamma_{twisted \,moduli}
&=& \frac{1}{12} \frac{1}{(4 \pi)^2} \sum_d \int d^4
\theta \left[ W^\alpha_d \right. W_{\alpha d}
\nonumber \\ &&
+ 3 \Tr_d (W^\alpha W_\alpha) + 3 (\Tr_d \left. W^\alpha)
W_{\alpha d} \right] \frac{1}{\Box} D^2 Z_d + h.c.\nonumber \\
\Gamma_{dilaton} &=& \frac{1}{12} \frac{1}{(4 \pi)^2} \int
d^4 \theta W^\alpha_S W_{\alpha S} \frac{1}{\Box} D^2 Z_S + h.c.
\eeqa 
The contribution from the gauginos has not been computed
in~[18].  However, it can be found in a similar way and
reads
\beqa             \Gamma_{gaugino} &=& \frac{1}{12} \frac{1}{(4
\pi)^2} \sum_H \int d^4 \theta \left[ n_V^H W^\alpha_V
\right. W_{\alpha V}  \nonumber \\
&+&  3 \Tr^H_V (W^\alpha W_\alpha) + 3 (\Tr^H_V \left. W^\alpha)
W_{\alpha V} \right] \frac{1}{\Box} D^2 Z_V + h.c. \eeqa 
where $H$ denotes the factor gauge groups $E_8$, $E_6$ and
$U(1)^2$, and where $n^H_V = \Tr^H_V \delta^a\,_b$.
Thus, the sum of all the non-local contributions from the light
fields to the effective action, as computed from
triangle graphs, is, to lowest order in prepotentials, given by
\beqa    \Gamma_{\rm light} &=& \Gamma_{VVZ} + \Gamma_{VZZ} +
\Gamma_{ZZZ}   \eeqa 
where
\beqa             \Gamma_{VVZ} &=& \frac{1}{4} \frac{1}{(4
\pi)^2} \int d^4 \theta \left[ \right. \sum_k \Tr_k (W^{\alpha}
W_{\alpha}) \frac{1}{\Box} D^2 Z_k \nonumber\\
&+& \left. \sum_d \Tr_d (W^{\alpha} W_{\alpha}) \frac{1}{\Box} D^2 Z_d
+ \Tr_V (W^{\alpha} W_{\alpha}) \frac{1}{\Box} D^2 Z_V   \right]
+ h.c. \eeqa  
and
\beqa    \Gamma_{VZZ} &=& \frac{1}{4} \; \frac{1}{(4 \pi)^2} \sum_H
\int d^4 \theta   \left[
\sum_k ( \Tr^H_k W^\alpha)
W_{\alpha k} \right. \frac{1}{\Box} D^2 Z_k  \nonumber\\
&+& \sum_d (\Tr^H_d W^\alpha) W_{\alpha d}
\frac{1}{\Box} D^2 Z_d
+ (\Tr^H_V W^\alpha)\left. W_{\alpha
V} \frac{1}{\Box} D^2 Z_V \right]  + h.c. \nonumber \\
\Gamma_{ZZZ} &=& \frac{1}{12} \frac{1}{(4 \pi)^2}
\int d^4 \theta
\left[ \sum_k n^{E_6}_k \right. W^\alpha _k W_{\alpha k}
\frac{1}{\Box} D^2 Z_k  \nonumber\\
&+& \sum_I W^\alpha _I W_{\alpha I}
\frac{1}{\Box} D^2 Z_I
+ \sum_d W^\alpha_d W_{\alpha d} \frac{1}{\Box}
D^2 Z_d  \nonumber\\
&+& \sum_H n_V^H W^\alpha_V W_{\alpha V} \frac{1}{\Box}
D^2 Z_V + W^\alpha_S W_{\alpha S} \left. \frac{1}{\Box} D^2 Z_S
\right]  + h.c. \eeqa
$\Gamma_{VVZ}$ contains all graphs of the form $vv(a+\Gamma)$
and, hence, it contains the subset of graphs $vv(a+\Gamma)$
previously discussed in~[3,7,8,9,10] and given by (and generalized
to the off-diagonal moduli case)
\beqa             \Gamma_{VVZ}^{sub} &=&
\frac{1}{4} \frac{1}{(4 \pi)^2} \sum_H \int d^4 \theta
(W^{(r) \alpha}_H W_{H (r) \alpha}) \frac{1}{\Box} D^2
\{ \left( \frac{1}{2} \kappa^2 (c^H_V - \sum_R c^H_R
) \right) K  \nonumber \\
&+& \sum_R c^H_R \lin \det {\cal G}_R \} + h.c. \eeqa 
where $R$ is any representation of charged chiral superfields,
$\Tr^H_{R,V}$ $T^{(r)} T^{(s)} = c^H_{R,V} \delta^{r s}$, and
where ${\cal G}_R$ is that portion of the \Kahler\ metric restricted to
chiral superfields with representation $R$ of factor group $H$.
It is this part of the effective action that is relevant
to the computation of non-harmonic gauge couplings.  The
contributions $\Gamma_{VZZ}$ and $\Gamma_{ZZZ}$ correspond to
all graphs of the form $v(a + \Gamma)^2$ and $(a + \Gamma)^3$,
respectively.  These terms have not previously been computed
and are one of the main results of this paper.  Note that
they include a pure \Kahler\ graph, $aaa$, a pure $\sigma$-model
coordinate graph, $\Gamma \Gamma \Gamma$, as well as new mixed
graphs such as $va\Gamma$ and $a{\Gamma}^2$.
Also note that mixed graphs such as $va\Gamma$ and $vaa$
contribute terms which are proportional to the trace of one
single Yang-Mills generator, $\Tr T^{(r)}$, which need not vanish
for the Yang-Mills gauge group under consideration, namely $E_8
\otimes E_6 \otimes U(1)^2$.
These graphs
render the theory anomalous under gauge transformations (4.2)
and, hence, are just as dangerous, and just as interesting as
the graphs contributing to $\Gamma_{VVZ}$.  It is tedious, and
not very enlightening, to rewrite all of $\Gamma_{VZZ}$ and
$\Gamma_{ZZZ}$ in terms of $K$, $\lin \det {\cal G}_{I \bar{J}}$
and $\lin \det {\cal G}_R$.  However, it is instructive to
rewrite the terms in $\Gamma_{ZZZ}$, which are proportional
to the square of the \Kahler\ fieldstrength $X_{\alpha}$
defined in (2.7), in this manner.  This
fieldstrength is contained in $W_k, W_I, W_d, W_V$ and $W_S$.
The $X^{\alpha} X_{\alpha}$ terms in $\Gamma_{ZZZ}$ (which
we generalize to the off-diagonal moduli case) can be
rewritten as
\beqa             \Gamma_{X^2} &=& \frac{1}{48} \frac{1}{(4
\pi)^2}  \int d^4 \theta  X^\alpha X_\alpha
\frac{1}{\Box}  D^2  [
\frac{1}{2} \kappa^2 K \left( - \sum_k (\prod_H n^H_k)
- \sum_d (\prod_H n^H_d) - n_I -1 \right. \nonumber\\
&+& \left. \sum_H n^H_V \right)
+ \lin \det {\cal G}_{I \bar{J}} + \lin {\cal G}_S
+ \sum_{sec} (\prod_H n^H_{sec})
\lin \det {\cal G}_{sec} ] + h.c. \eeqa 
where $n_{k,d}^H = \Tr_{k,d}^H \delta_b^a$,
$n_I$ denotes the number of untwisted moduli, $\sum_{sec}$
implies a sum over all sectors of matter and twisted moduli, $
{\cal G}_{sec}$ is that portion of the \Kahler\ metric restricted
to the sector $sec$, and where $(\prod_H n_{sec}^H)$ counts the
Yang-Mills degrees of freedom in each of the sectors.

Let us discuss the anomalous behaviour of the various
parts of $\Gamma_{\rm light}$ under \Kahler\ and
$\sigma$-model coordinate
transformations.  To begin with, note that
$K$, $\lin \det {\cal G}_{I
\bar{J}}$ and $\lin \det {\cal G}_R$, which are contained in the various
superfields $Z$, are very different geometrical
objects, as made explicit by the different powers of $\kappa^2$
multiplying them.  As such, they will in general
transform very differently under \Kahler\ transformations and
under general coordinate transformations on the moduli and matter
manifold.  Under \kahler\ transformations, only $K$ transforms
whereas $\lin \det {\cal G}_{I \bar{J}}$ and $\lin \det
{\cal G}_R$ are inert.  Under general coordinate transformations,
and, hence, under the subset of isometries, on the
submanifold of the untwisted moduli,
$K$ and $\lin \det {\cal G}_R$ transform as scalar functions
\beqa K'(T', \bar{T}', S, \bar{S}) &=& K(T,\bar{T}, S, \bar{S})
\nonumber \\
\lin \det {\cal G}'_{i \bar{j}}(T', \bar{T}') &=& \lin \det {\cal G}_
{i \bar{j}}(T, \bar{T})   \eeqa
whereas $\lin \det {\cal G}_{I \bar{J}}$ transforms as the log of
a density
\beqa \lin \det {\cal G}'_{I' \bar{J}'}(T', \bar{T'}) &=&
\lin \det {\cal G}_{I \bar{J}}(T, \bar{T}) + \lin \det \Lambda_{I'}\,
^{I} + \lin \det \bar{\Lambda}_{\bar{J}'}\,^{\bar{J}}  \eeqa 
where
\beq {\Lambda}_{I'}\,^I = \frac{\partial T^I}{\partial
T'^{I'}}  \eeq 
Similarly, under coordinate transformations on the submanifold of
matter fields with metric ${\cal G}_R$, the quantities $K$ and
$\lin \det {\cal G}_{I \bar{J}}$ transform as scalar functions
whereas
$\lin \det {\cal G}_R$ transforms
as the log of a density.
First, let us apply these results to compute the variation
of the various parts of $\Gamma_{\rm light}$ under general
\kahler\ transformations
\beq K \ra K + F (T,S) + \bar{F} (\bar{T}, \bar{S}) \eeq 
As an example, it is easy to see from (4.20) that
under \Kahler\ transformations (4.25)
$\Gamma_{VVZ}^{sub}$ transforms as
\beqa        \delta_K \Gamma_{VVZ}^{sub}
= - \frac{1}{2} \frac{1}{(4 \pi)^2} \sum_H \int d^2 \theta  \kappa^2
F(T,S) (W^{(r) \alpha}_H W_{H (r) \alpha})
(c^H_V - \sum_R c^H_R)     + h.c. \eeqa  
The right hand side of this expression is non-zero and, hence,
$\Gamma_{VVZ}$ is \Kahler\ anomalous.
In a similar way, one can show that $\Gamma_{VZZ}$ and
$\Gamma_{ZZZ}$ are anomalous as well under
\Kahler\ transformations.  For example, it follows from
(4.21) that $\Gamma_{X^2}$ transforms as
\beqa             \delta_K \Gamma_{X^2} &=& \frac{1}{24}
\frac{1}{(4 \pi)^2} \int d^2
\theta \kappa^2 F(T,S) X^\alpha X_\alpha[
\sum_k (\prod_H n^H_k) + \sum_d (\prod_H n^H_d) + n_I \nonumber\\
&+& 1 - \sum_H n^H_V ] + h.c.  \eeqa 
Again, the right hand side of this expression doesn't
vanish and, hence, $\Gamma_{X^2}$ is \Kahler\ anomalous.
Note that $\delta_K \Gamma_{X^2}$ arises from a
triangle graph in which all
three vertices couple to $K$ only.  Thus, it corresponds to a
pure \Kahler\ anomaly.
Now let us consider
the behaviour of the various parts of $\Gamma_{\rm light}$
under $\sigma$-model coordinate transformations.  It follows
from (4.23) that $\Gamma_{X^2}$, given in (4.21), does
not transform as a scalar under $\sigma$-model coordinate
transformations on the submanifold of the untwisted moduli.
Similarly, $\Gamma^{sub}_{VVZ}$ given in (4.20) does not
transform as a scalar under $\sigma$-model coordinate
transformations on the submanifold of the matter fields.
That is, the one-loop effective action $\Gamma_{\rm light}$
does not transform as a scalar under $\sigma$-model coordinate
transformations and, hence, the theory is anomalous under
these transformations.  Finally, let us consider the behaviour of the
theory under a particular subset of $\sigma$-model coordinate
transformations, namely under isometries (3.9) on the submanifold
of the untwisted moduli.
First consider $\Gamma_{VVZ}^{sub}$ given in (4.20).  We
will choose the standard $(T,S)$-coordinate
system in which $K$ takes the form (4.1).  In this
coordinate system, both $\lin \det {\cal G}_{I \bar{J}}$ and $\lin \det
{\cal G}_R$ are proportional to $K - K_S$.  $K$, however, transforms as
in (3.13) under isometries (3.9), and, hence, $\Gamma_{VVZ}^{sub}$
transforms as
\beqa   \delta    \Gamma_{VVZ}^{sub}
= - \frac{1}{2} \frac{1}{(4 \pi)^2} \sum_H \sum_I
\int d^2 \theta
F^I (W^{(r) \alpha}_H W_{H (r) \alpha})
(c^H_V - \sum_R c^H_R (1 - 2 q^I_R))+ h.c. \eeqa 
That is, $\Gamma_{VVZ}$ is anomalous under these isometries.
Non-invariance of $\Gamma_{VVZ}^{sub}$ under modular
transformations,
that is, isometries (3.9) for which $a$, $b$, $c$, $d \in \Z$,
follows from this result and has
been extensively discussed~[3,8,9] in the context of
non-harmonic gauge couplings in string theory.  In a similar way,
one can see that the new contributions $\Gamma_{VZZ}$ and
$\Gamma_{ZZZ}$ are anomalous under $\sigma$-model isometries.
For example, it follows from (4.21) that $\Gamma_{X^2}$ transforms
as
\beqa             \delta \Gamma_{X^2} &=& - \frac{1}{24}
\frac{1}{(4 \pi)^2} \sum_I \int d^2
\theta F^I X^\alpha X_\alpha[
- \sum_k (\prod_H n^H_k) - \sum_d (\prod_H n^H_d) - n_I - 1 \nonumber\\
&+& \sum_H n^H_V +4 + 2 \sum_{sec} ( \prod_H n^H_{sec}) q^I_{sec}]
+ h.c.  \eeqa 
and, hence, is non-invariant under isometries (3.9) (as well
as under the subset of modular transformations).

We would like to point out again that, in the entire
discussion given above, we have not included matter fields
which are singlets under the $E_6$, since the structure
of their \Kahler\ potentials is unknown.  They do, however,
in general contribute to $\Gamma_{\rm light}$ as well.  Hence, one
might ask whether their inclusion results in a cancellation
of the various anomalous parts in $\Gamma_{\rm light}$.  We show,
in Appendix A, that this does not happen.  For concreteness,
we consider in Appendix A the case of the $Z_3$ orbifold, for which
the complete \Kahler\ potential is known, and we show that
$\Gamma_{X^2}$, and, hence, $\Gamma_{\rm light}$, is non-vanishing.

Finally, it should also be pointed out that in the entire discussion
given above, we have ignored the fact that the gravitino couples
to the $U_K(1)$ connection as well.  That is, we have avoided the
issue of the quantization of supergravity fields.  One might ask,
however, whether the inclusion of the contribution of the
gravitino field in (4.17) could result in a cancellation of the
pure \kahler\ anomaly.  From the appropriate
index-theorem~[19] for the spin $\frac{3}{2}$ operator of a
Rarita-Schwinger spinor $\psi_m ^\alpha$, it follows that the
non-local contribution of the gravitino to $\Gamma_{X^2}$ is three times
as big as the contribution of a single spin $\frac{1}{2}$
fermion with prepotential superfield $Z$ given by $Z =
\frac{\kappa^2}{2} K$.
Hence, the gravitino field contributes a factor -3 to the
anomaly coefficient in (4.27).
That is, the pure \Kahler\ anomaly remains non-vanishing.

\section{Cancellation of K\"{a}hler and $\bf \sigma$-model
Coordinate Anomalies}

\setcounter{equation}{0}

\hspace*{.3in} In this section we will propose a cancellation
mechanism for the \kahler\ and $\sigma$-model coordinate
anomalies
in the effective theory of light fields, as computed in the
previous section.  We will proceed as follows.  We will demand
the theory to be \kahler\ invariant as well as
invariant under $\sigma$-model coordinate transformations.
To achieve this, we
will have to postulate the existence of two types of
counterterms, one for each type of symmetry transformation.
These counterterms, when added to the effective \lag\ of the
light fields, will have to restore both types of symmetries.  The
presence of these counterterms will be interpreted as resulting
from integrating out massive fields, such as Kaluza-Klein
excitations and winding modes.  We will be able to determine
the moduli dependence
of the two types of counterterms in the
standard ($T,S$)-coordinate system from the requirement of
modular invariance of the effective theory.  Finally, we introduce
an extension of the $\sigma$-model Green-Schwarz mechanism which
will enable us to remove the universal part of $\Gamma_{X^2}$,
as well as the universal part of the $\Gamma_{VVZ}$ term~[10]
discussed in the context of non-holomorphic gauge couplings in
string theory.  As in the previous sections, we will focus on
the $ E_8 \otimes E_6 \otimes U(1)^2$ orbifolds.  We conjecture,
however, that for other (2,2) symmetric orbifolds, such as $Z_3$,
all the steps carried out in our analysis remain valid.  For
concreteness, some explicit results are given in Appendix A
for the case of the $Z_3$ orbifold.

We begin by demanding \kahler\ invariance of the effective theory
of (2,2) symmetric orbifolds.  As pointed out in the last
section, the effective \lag\ $\Gamma_{\rm light}$ is not
invariant under \kahler\ transformations (4.25).  Hence, a first
type of counterterm has to be introduced.  One way of rendering
the effective theory \kahler\ invariant is to conjecture that the
integration over massive modes conspires in such a way as to
effectively induce a shift of the \kahler\ potential (4.1) in
$\Gamma_{\rm light}$ (contained in the $Z$'s) by an amount
\beq \kappa^2 K \ra \kappa^2 K + \ln |W|^2 \equiv \kappa^2 G(T,
\bar{T}, S, \bar{S}) \eeq 
Here, $W(T,S)$ denotes a yet unspecified superpotential.  We have
adopted a minimalestic principle.  That is to say, we assume each
term proportional to $K$ in (4.17) gets shifted by the same
amount $\ln |W|^2$.

Next, we would like to demand invariance of the partition function
of (2,2) symmetric orbifolds under $\sigma$-model coordinate
transformations.  We will proceed in two steps.  In the first
step we will demand the effective Lagrangian to transform
as a scalar under general $\sigma$-model coordinate transformations.
As pointed out in the
last section, the effective \lag\ $\Gamma_{\rm light}$, computed
from the light fields, does not transform as a scalar function
under general coordinate transformations on the moduli, matter
and dilaton manifold.  This is so, because the contributions
proportional to $Z_A$ (where $A = k, I, d, S)$ contain $\lng_A$
terms.  A term $\lng_A$, however, transforms as the log of a
density under
an $A$-type $\sigma$-model coordinate transformation.  Hence, a
second set of counterterms, denoted $P_A$, has to be introduced.
One way of achieving the scalar behaviour under $A$-type
$\sigma$-model coordinate transformations is to conjecture that
the integration over heavy modes contributes in such a way as to
effectively shift the $\lng_A$ prepotentials in $\Gamma_{\rm
light}$ by amounts
\beq \lng_A \ra \lng_A - P_A \eeq 
where
\beq P_A = \ln \hat{\G}_A \eeq 
and
\beq \hat{\G}_A = \partial_A \partial_{\bar{A}} P \eeq 
$P$ denotes some yet unspecified scalar function of the moduli,
matter and dilaton fields.  Then, $P_A$ clearly transforms as the log
of a density under $A$-type $\sigma$-model coordinate transformations
and, hence, the combination (5.2) transforms as a scalar function
under any $\sigma$-model coordinate transformation.
Note that the combination $\lng_A - P_A$ is \kahler\ invariant by
construction.

We now demand that the effective theory be invariant
under the $SL(2, \Z)$ modular transformations, to
conform to the expectation that superstring amplitudes
are modular invariant~[11].
Therefore, we must demand that the
scalars $\lng_A - P_A$ and $G(T, \bar{T}, S, \bar{S})$ be
invariant under the modular transformations
of the untwisted moduli.  We will choose the standard
$(T,S)$-coordinate system, in which the \kahler\ potential $K$ is
given by (4.1).  In particular, then,
by demanding the scalars
$\lng_A -P_A$ and $G(T, \bar{T}, S, \bar{S})$ to be separately
invariant under modular transformations,
we will be able to find the
explicit form of some of the counterterms in the standard
$(T,S)$-coordinate system.  First, let us
discuss $G(T, \bar{T}, S, \bar{S})$.  If we make the ansatz
\beq W (T,S) = \Omega (S) W(T) \eeq 
then
\beq G(T, \bar{T}, S, \bar{S}) = G(T,\bar{T}) + G(S, \bar{S})
\eeq 
where
\beqa G(T, \bar{T}) &=& K_M (T, \bar{T}) + \kappa^{-2} \ln
|W(T)|^2 \nonumber \\
 G(S, \bar{S}) &=& K_S (S, \bar{S}) + \kappa^{-2} \ln
|\Omega(S)|^2 \eeqa 
Demanding the invariance of $G(T, \bar{T}, S, \bar{S})$ under
modular transformations, the $SL(2, \Z)$ subset of isometries
(3.9), restricts $W(T)$ to be the modular function
\beq W(T) = \prod_I \eta^{-2} (i T^I) \eeq 
It follows that
\beq \kappa^2 G(T, \bar{T}) = - \sum_I \ln (T^I + \bar{T}^I) |
\eta (iT^I)|^4 \eeq 
The explicit form of $W(T)$, and, hence, of $G(T,\bar{T})$ agrees
with recent discussions in the literature~[9].  Note that
$\Omega(S)$ remains unspecified.  Next, let us discuss the
counterterms $P_A$, $A = k, d$, for matter fields and
twisted moduli.  It follows from (3.2) that $\lng_A$ is given as
\beq \lng_A = \sum_I \left( -q^I_A \ln \left( T^I + \bar{T}^I
\right) \right) \eeq 
Hence, the scalar $\lng_A - P_A$ is invariant under the
modular transformations provided that
\beq \hat{\G}_A = \prod_I \left( \left| \eta(iT^I)\right|^4
\right)^{q^I_A} \eeq 
The scalars $\lng_A - \ln \hat{\G}_A$, $A= k, d$, have the
same form as the one-loop corrections to inverse gauge couplings
in string theory computed in~[7,9]. Similarly, the counterterm
$\hat{\G}_I$ for untwisted moduli is determined to be
\beq \hat{\G}_I = \prod_I \left( \left| \eta (iT^I) \right|^4
\right)^2 \eeq 
Note that the counterterm $P_S$ remains unspecified.
Although the counterterms (5.8), (5.11) and (5.12)
restore invariance of the effective Lagrangian under
$SL(2, \Z)$ transformations, they do not restore the full
$SL(2, \R)$ isometry invariance of the tree-level kinetic
Lagrangian (3.6).  That is, the $SL(2, \R)$ symmetry of
the tree-level kinetic Lagrangian is broken down to
$SL(2, \Z)$ modular invariance at the one-loop level.

We have rendered the theory of (2,2) symmetric
orbifolds invariant under \kahler\ transformations and under
general $\sigma$-model coordinate transformations
by introducing two types of
counterterms, one for each type of symmetry transformation. A
priori, there is then no need for any additional mechanism for
restoring the symmetries under consideration.  However,
it has been shown~[10] in explicit calculations of string theory
scattering amplitudes that there is yet another removal mechanism
at work, namely a $\sigma$-model generalization~[8] of the
Green-Schwarz mechanism~[12].  Hence, we now proceed to
discuss extensions of the Green-Schwarz mechanism for any
(2,2) symmetric orbifold~[8].  The Green-Schwarz term, denoted
$\L_{GS}$, utilizes the linear multiplet representation $l$ for
the dilaton supermultiplet~[20].  The linear multiplet $l$ is
defined to satisfy the following modified linearity condition in
\kahler\ superspace
\beqa        (\bar{\D}^2 - 8R) l &=& - 8 \sum_H t_H (W_H^{(r) \alpha}
W^{(r)}_{H\alpha})  + \frac{1}{6} (3\sigma + 4\tau)X^\alpha X_\alpha
\nonumber \\
&& - 8\tau (W^{\alpha \beta \gamma} W_{\alpha \beta \gamma}) \nonumber \\
(\D^2 - 8R^+) l \ &=& - 8 \sum_H t_H (\bar{W}^{(r)}_{H\dot{\alpha}}
\bar{W}^{(r)\dot{\alpha}}_H) + \frac{1}{6} (3 \sigma + 4\tau)
\bar{X}_{\dot{\alpha}} \bar{X}^{\dot{\alpha}} \nonumber \\
&& - 8\tau (\bar{W}_{\dot{\alpha} \dot{\beta} \dot{\gamma}}
W^{\dot{\alpha} \dot{\beta} \dot{\gamma}}) 
\eeqa
The $X^\alpha X_\alpha$ terms in (5.13) correspond to a
Chern-Simons term for the \kahler\ symmetry and are a
generalization of the $U_R(1)$ Chern-Simons terms introduced
in~[21].
Equation (5.13) describes the most general coupling of a linear
multiplet to the square of gauge, \Kahler\ and gravitational superfield
fieldstrengths.  That is, in a general supergravity theory the
coefficients $t_H$, $\sigma$ and $\tau$ in (5.13) can be taken to
have arbitrary values.  In the case of the superstring inspired
supergravity theories we are mainly concerned with, these
coefficients have fixed values.  For instance, one finds that, if
the description of the dilaton supermultiplet in the chiral $S$
picture is to be equivalent to its description in the $l$
picture, each of the Yang-Mills coefficients $t_H$ has to be
taken to satisfy
\beq t_H =  \frac{1}{24} \eeq 
Then, it can be
          shown~[20] that, upon duality
transformation from the chiral superfield $S$ to $l$, the
tree-level \lag\ (3.6) in the chiral $S$ picture is equivalent
to the following \lag\ in the $l$ picture
\beq \L = - 3 \kappa^{-2} \int d^4 \theta \frac{2}{3} E \eeq 
where $E = E [\hat{K}]$ with $\hat{K} = K_M + \ln l$, and where
$l$ satisfies (5.13) with coefficients $t_H$ given by (5.14).
Equivalence of tree-level \lag s (3.6) and (5.15) also implies
that the coefficients $\sigma$ and $\tau$ vanish.  In the
following, however, we will allow for non-vanishing $\sigma$ and
$\tau$ and, hence, we will discuss extensions of the supergravity
theories given by (3.6).  That is, we will allow for additional
couplings of the dilaton $S$ to the square of \kahler\ and
gravitational superfield field strengths.  In fact, by
generalizing the duality transformation given in [20] it can be
shown that the \kahler\ and gravitational contributions on the
right hand side of (5.13) contribute the following additional
terms to the tree-level \lag\ (3.6) in the $S$-picture
\beqa \L' &=& - \frac{(3 \sigma + 4 \tau)}{16}
 \int d^4 \theta \frac{E}{R} S
X^\alpha X_\alpha +
3 \tau \int d^4 \theta \frac{E}{R} S W^{\alpha \beta \gamma}
W_{\alpha \beta \gamma} + h.c. \eeqa 
We will determine the coefficients $\sigma$ and $\tau$ in the
following way.  We will conjecture that a part (perhaps all)
of the universal part of the
mixed $\sigma$-model gravitational anomalies, as discussed in [3]
and in Appendix~A, as well as a part (perhaps all) of the
pure \Kahler\ anomaly get removed by an extension of the
Green-Schwarz mechanism.  This will uniquely fix both
coefficients $\sigma$ and $\tau$.  Hence, removal of these
universal parts in string theory requires the presence of the
additional tree-level \kahler\ and gravitational couplings
(5.16).

We proceed to discuss the $\sigma$-model generalization of the
Green-Schwarz mechanism.  We first expand (5.13) to lowest order
in the prepotentials $V$ and $K$, and we find that
\beqa             \bar{D}^2 l &=& - \frac{1}{3} \sum_H
(W^{(r)\alpha}_H W^{(r)}_{H \; \alpha}) + \frac{1}{6} (3 \sigma +
4 \tau) X^\alpha X_\alpha \nonumber \\
                 {D}^2 l &=& - \frac{1}{3} \sum_H
(\bar{W}^{(r)}_{H\dot{\alpha}} \bar{W}^{(r)\dot{\alpha}}_{H
}) + \frac{1}{6} (3 \sigma + 4 \tau) \bar{X}_{\dot{\alpha}}
\bar{X}^{\dot{\alpha}} \eeqa 
Note that, to lowest order in the prepotentials $V$ and $K$, the
superfield $W_{\alpha \beta \gamma}$ doesn't contribute to the
right hand side of equation (5.17).  Also note
that, on the right hand side of (5.17), there is no superfield
fieldstrength's square associated with prepotentials
ln ${\cal G}_A$. The Green-Schwarz term, which we will take to
be of order $\hbar$, has the generic form
\beq \L_{GS} = \beta \int d^4 \theta E l \Xi + h.c. \eeq 
where $\beta$ is real and where $\Xi$ is taken to be a scalar
function of the untwisted moduli $T^I$, invariant under
\kahler\ transformations and under modular transformations.
Hence, the addition of the term $\L_{GS}$ to the \lag\ of
the effective theory (already supplemented with appropriate
counterterms, as discussed above) will maintain both \kahler\ and
modular invariances of the theory.  The
coefficient $\beta$ as well as the
scalar function $\Xi$ in (5.18) can be determined
from the knowledge (in any (2,2) symmetric $Z_N$ orbifold)
of the absence
of dependence of the effective $E_8$ gauge coupling
constant on the untwisted moduli~[10].  We proceed as follows.
We first solve (5.17)
for $l$
\beqa l &=& \frac{1}{16} \; \frac{D^2}{\Box} \left\{ - \frac{1}{3}
\right. \sum_H \left( W^{(r)\alpha}_H W^{(r)}_{H\;\;\alpha}
\right) +
 \left. \frac{1}{6} \left( 3 \sigma + 4\tau \right) X^\alpha
X_\alpha \right\}  \nonumber \\
&+& \frac{1}{16} \; \frac{\bar{D}^2}{\Box} \left\{ - \frac{1}{3}
\right. \sum_H \left( \bar{W}_H^{(r)} \,_{\dot{\alpha}}
\bar{W}_H^{(r) \dot{\alpha}} \right) +
 \left. \frac{1}{6}      ( 3 \sigma + 4 \tau)
\bar{X}_{\dot{\alpha}} \bar{X}^{\dot{\alpha}} \right\} \eeqa
Then, inserting (5.19) into (5.18) yields the following non-local
contribution $\Gamma_{GS}$ to the effective theory
\beqa             \Gamma_{GS} &=& - \frac{1}{24} \beta \sum_H \int
d^4 \theta (W^{(r) \alpha}_H W^{(r)}_{H\alpha}) \frac{1}{\Box}
D^2 \Xi \nonumber\\
&+& \frac{1}{8} \beta \frac{(3\sigma + 4\tau)}{6} \int d^4 \theta
X^\alpha X_\alpha \frac{1}{\Box} D^2 \Xi + h.c. 
\eeqa
The two terms in $\Gamma_{GS}$ stem from reducible tree
graphs similar to the one shown in Figure 2.
We will call them contributions of the universal type.
We now compare $\Gamma_{GS}$ in (5.20) against $\Gamma_{VVZ}^{sub}$
in (4.20).  We supplement $\Gamma_{VVZ}^{sub}$ with counterterms
(5.1) and (5.4) and we denote the resulting expression by
${\Gamma}_{VVZ}^{'sub}$.  It is useful to rewrite ${\Gamma}_{VVZ}^{'sub}$
in the following way
\beqa             \Gamma_{VVZ}^{'sub} &=&
\frac{1}{4} \frac{1}{(4 \pi)^2} \sum_H \int d^4 \theta
(W^{(r) \alpha}_H W_{H (r) \alpha}) \frac{1}{\Box} D^2
\left( \frac{1}{2} \kappa^2 c^{E_8}_V  \right) G  \nonumber \\
&+& \frac{1}{4} \frac{1}{(4 \pi)^2} \sum_H \int d^4 \theta
(W^{(r) \alpha}_H W_{H (r) \alpha}) \frac{1}{\Box} D^2
\{ \left( \frac{1}{2} \kappa^2 ( - c^{E_8}_V + c^H_V - \sum_R c^H_R
) \right) G  \nonumber \\
&+& \sum_R c^H_R \left( \lin \det {\cal G}_R - \lin \det
\hat{\G}_R  \right) \} + h.c. \eeqa 
The first sum on the right hand side of $\Gamma_{VVZ}^{'sub}$
yields a moduli dependent contribution to the effective $E_8$
gauge coupling constant.  It has been shown~[10], however, that
such a contribution is absent in the effective theory of any
(2,2) symmetric $Z_N$ orbifold.  Hence, we will remove this
contribution in the following way.  By noting that the term
$\kappa^2 c_V^{E_8} G(T, \bar{T})$ in
$\Gamma_{VVZ}^{'sub}$ is of the universal type and by comparing
$\Gamma_{VVZ}^{'sub}$ against $\Gamma_{GS}$ in (5.20), it
follows that the term $\kappa^2 c_V^{E_8} G(T, \bar{T})$
can be removed from $\Gamma_{VVZ}^{'sub}$ by choosing
\beqa \Xi = \kappa^2 G(T, \bar{T}) \eeqa 
and
\beq \beta = 3 \; \frac{1}{(4 \pi)^2} c_V^{E_8} \eeq
where, in the standard coordinate system, $G(T, \bar{T})$
is given by (5.9).
$\Gamma_{GS}$ now becomes
\beqa \Gamma'_{GS}& =& \frac{1}{8} \beta \frac{(3 \sigma + 4
\tau)}{6} \int d^4 \theta X^\alpha X_\alpha \frac{1}{\Box} D^2
\kappa^2 G(T, \bar{T}) + h.c. \eeqa 
We now compare $\Gamma'_{GS}$ against $\Gamma_{X^2}$ in
(4.21).  We supplement $\Gamma_{X^2}$ with counterterms
(5.1) and (5.4) and we denote the resulting expression
by $\Gamma'_{X^2}$.  We rewrite $\Gamma'_{X^2}$ as follows
\beqa             \Gamma'_{X^2} &=& \frac{\alpha}{96}
\frac{1}{(4 \pi)^2}  \int d^4 \theta  X^\alpha X_\alpha
\frac{1}{\Box}  D^2 \kappa^2 G \nonumber \\
&+&  \frac{1}{48} \frac{1}{(4
\pi)^2}  \int d^4 \theta  X^\alpha X_\alpha
\frac{1}{\Box}  D^2  [
\frac{1}{2} \kappa^2 G \left( - \alpha - \sum_k (\prod_H n^H_k)
- \sum_d (\prod_H n^H_d) \right. \nonumber\\
&-&  \left. n_I -1 + \sum_H n^H_V \right)
+ \lin \det {\cal G}_{I \bar{J}} - \lin \det \hat{\G}_{I
\bar{J}} + \lin {\cal G}_S  - \lin \hat{\G}_S \nonumber \\
&+& \sum_{sec} (\prod_H n^H_{sec}) \left(
\lin \det {\cal G}_{sec} - \lin \det \hat{\G}_{sec} \right)]
+ h.c. \eeqa 
where $\alpha$ denotes an yet unspecified constant.
Note that $\Gamma'_{X^2}$
contains terms which are of the universal type (5.24).  These
terms stem from the pure \Kahler\ graph $aaa$.  We now
conjecture that a certain amount of the universal
type gets removed by the Green-Schwarz mechanism as well.  We
will denote the amount by $\alpha$.
The actual value of $\alpha$ has yet
to be determined by an explicit calculation of the
appropriate three-point string scattering amplitude.  Here, we
assume that $\alpha$ is non-vanishing.
By comparing $\Gamma'_{X^2}$ against $\Gamma'_{GS}$, it
follows that the first term in $\Gamma'_{X^2}$ can
be removed by choosing
\beq 3 \sigma + 4 \tau = - \alpha \;
\frac{1}{6 c_V^{E_8}}  \eeq 
We would like to point out again, that we have omitted
the contributions from matter singlets to $\Gamma'_{X^2}$,
since the structure of their \Kahler\ potentials is
unknown.  For concreteness, we will, in Appendix A,
discuss the Green-Schwarz mechanism for the case of
the $Z_3$ orbifold, for which the complete \Kahler\
potential is known.
Let us also emphasize again that, whether or not the removal
(and how much) of a universal piece of $\Gamma'_{X^2}$
by the Green-Schwarz mechanism takes place in the effective
theory, has yet to be checked by an explicit calculation of the
appropriate three-point scattering amplitude.

We have, so far, only determined the linear
combination $3 \sigma + 4 \tau$ of coefficients $\tau$ and
$\sigma$.  We will now determine $\tau$ and $\sigma$
individually, as follows.
The coefficient $\tau$ can be determined by conjecturing that
an amount $\gamma$
of the anomalous mixed $\sigma$-model gravitational
1-loop contributions to $\Gamma_{\rm light}$ is removed by the
Green-Schwarz mechanism as well.  Such contributions arise from
triangle graphs $\omega \omega a$ and $\omega \omega \Gamma$,
where $\omega$ denotes the space-time Lorentz-connection.  We
have not discussed such graphs in this paper; they have, however,
been discussed in [3].  In Appendix A we will compute these mixed
gravitational contributions for the case of the $Z_3$ orbifold,
as an example.  We also include the contribution from the
gravitino~[22], which had been missing in [3].  Similarly to
the $Z_3$ case, the contributions of the universal type
in $U(1)^2$ orbifolds arise in the graph $\omega \omega a$.
These contributions to $\Gamma_{\rm light}$ are proportional
to $K$ and have the form
\beqa \Gamma_{\rm gravity}^{univ} &=& \frac{1}{12} \;
\frac{1}{(4\pi)^2}
\int d^4 \theta W^{\alpha \beta \gamma} W_{\alpha \beta \gamma}
\frac{1}{\Box} D^2 (\gamma + (c - \gamma)) \kappa^2 K
+ h.c. \eeqa  
where the coefficient $c$ can be determined from the number
of massless states in the orbifold spectrum, and where
$\gamma$ denotes the amount of universal type we conjecture
is removed by the Green-Schwarz mechanism.
The removal of the
amount $\gamma$ fixes the coefficient
$\tau$ to be equal to
\beq \tau = \gamma \; \frac{1}{36 c_V^{E_8}} \eeq 
Then, the coefficient $\sigma$ is determined from (5.26)
and (5.28)  to be equal to
\beq \sigma = - (\frac{\gamma}{3} + \frac{\alpha}{2}) \;
\frac{1}{9 c_V^{E_8}}   \eeq 
The non-vanishing of the coefficients $\tau$ and $\sigma$
leads to the addition of the
new contributions (5.16) to the tree-level
Lagrangian (3.6) in the S-picture.  For concretenes,
we would like to display the bosonic component
content of the term in (5.16) proportional to the square of
the \Kahler\ superfield fieldstrength
$X^2$
\beqa {\cal L}'_{X^2} &=& - \frac{1}{8} ( 3\sigma + 4\tau )
\kappa^4 S| ( 2 v_{mn}^K v^{Kmn} + i
\epsilon^{klmn} v_{kl}^K v_{mn}^K + (D^{K})^2 ) + h.c.    \eeqa 
where $ \left. S \right|$ denotes the lowest component of
the dilaton superfield $S$,
where $v_{mn}^K$ denotes the fieldstrength of the
component \Kahler\ connection (3.15), $v_{mn}^K = \partial_m
a_n - \partial_n a_m$, and where $D^K$ is given in (4.12).
 Finally,
let us point out again that, in the above discussion, we have not
been able to fix the form of the counterterms $\Omega (S)$ and
$P_S$.

\section{Conclusion}

\hspace*{.3in} We have calculated and discussed the complete
set of one-loop triangle graphs involving the Yang-Mills
gauge connection, the \Kahler\ connection and the $\sigma$-model
coordinate connection in the effective field theory of (2,2)
symmetric $Z_N$ orbifolds.  We have shown that each of these
graphs is anomalous under both \Kahler\ transformations and
$\sigma$-model coordinate transformations.
We have demonstrated that, upon
introducing certain counterterms, it is
possible to preserve both
\kahler\ and $\sigma$-model coordinate invariances
of $Z_N$ orbifolds.  We have
been able to determine the moduli dependence
of these counterterms by demanding $SL(2, \Z)$ modular
invariance of the effective theory of $Z_N$ orbifolds.
The $SL(2, \R)$ symmetry of the tree-level kinetic Lagrangian
is broken by the counterterms down to $SL(2, \Z)$ modular
invariance. Additional
explicit calculations of on-shell three-point correlation
functions (involving the untwisted moduli and the dilaton) are
necessary to determine the dilaton dependence
of these counterterms.  We have
also discussed the possibility of removing the universal
part of the contribution from the pure
\kahler\ graph $aaa$
by the Green-Schwarz mechanism.  Whether or not such a
removal takes place in the effective theory has yet to be checked
by an explicit calculation of the appropriate composite
three-point scattering amplitude.  The relevant on-shell
three-point correlation function consists of two \kahler\ gauge
fields $a_m$, $a_n$ and one untwisted modulus $T$.
We would like to emphasize the importance of
such calculations and to urge people to do them.

\section*{Acknowledgements}

\hspace*{.3in} We would like to thank M. Cveti\v{c} for many
informative
discussions.  We are grateful to R. Grimm for many fruitful
discussions on the geometry of \kahler\ superspace.

\section*{Appendix A}

\setcounter{equation}{0}
\renewcommand{\theequation}{A.\arabic{equation}}

\hspace*{.3in} We give a detailed discussion of the
anomalous mixed $\sigma$-model gravitational 1-loop
contributions as well as of
the $\Gamma_{X^2}$ contribution to $\Gamma_{\rm light}$
for the case of the
$Z_3$ orbifold.  We begin with
the anomalous mixed $\sigma$-model
gravitational 1-loop contributions.  Such
contributions arise from triangle graphs $\omega \omega a$ and
$\omega \omega \Gamma$, where $\omega$ denotes the space-time
Lorentz connection.  They are invariant under space-time
Lorentz transformations, since, as it is well known~[19],
there are no pure space-time gravitational anomalies in
four dimensions. The anomalous mixed $\sigma$-model
gravitational 1-loop contributions were discussed in [3]
and found to be equal to
\beqa \Gamma_{\rm gravity} &=& \frac{1}{12} \;
\frac{1}{(4\pi)^2}
\int d^4 \theta W^{\alpha \beta \gamma} W_{\alpha \beta \gamma}
\frac{1}{\Box} D^2 (c \kappa^2 K + N) + h.c. \eeqa  
where
\beq c = 720 \eeq 
and
\beqa N &=& \kappa^2 K ( - c - \sum_k ( \prod_H n_k^H )
- \sum_d ( \prod_H n_d^H ) - n_I - 1 + \sum_H n_V^H -21 )  \nonumber\\
&+& 2 \ln \det {\cal G}_{I \bar{J}} + 2 \ln {\cal G}_S
+ 2 \sum_{sec} ( \prod_H
n_{sec}^H ) \ln \det \G_{sec}  \eeqa 
The quantities on the right hand side of equation (A.3) have been
defined in section 4 of this paper.
In (A.3), we have included the contribution from the dilatino, $-1$,
as well as from the gravitino, $-21$.  The latter contribution
was computed
in [22] and can be readily obtained from the appropriate
index-theorem~[19] for the spin $\frac{3}{2}$ operator of a
Rarita-Schwinger spinor.
We would like to compute (A.3) for the $Z_3$
case and show that, in the standard $(T,S)$-coordinate system,
the dependence of $N$ on the untwisted moduli cancels out.  We
begin by computing the coefficient of the $K$-term in (A.3).  The
gauge group $G$ of $Z_3$, $G=E_8 \otimes E_6 \otimes SU(3)$,
contains $\sum_H n_V^H = 334$ generators.
Massless chiral superfields
occur in both the untwisted and twisted sectors.  In the
untwisted sector, there are nine (1,1) moduli superfields,
denoted $T_{I\bar{J}}$ where $I, \bar{J} = 1,2,3$, which
transform as {\bf (1,1,1)} under $G$.  In addition there are three
matter superfields which transform as {\bf (1,27,3)} under $G$.  In
the twisted sector there are twenty seven matter superfields
which transform as {\bf (1,27,1)} under $G$ as well as eighty-one
superfields which transform as {\bf (1,1,${\bf \bar{3}}$)}
under $G$.
Inserting this information
into (A.3) yields
\beqa  - c - \sum_k ( \prod_H n_k^H )
- \sum_d ( \prod_H n_d^H ) - n_I - 1 + \sum_H n_V^H -21  = -1632
\eeqa 
Next, we compute the terms proportional to $\ln \det \G$ in
(A.3).  This requires the knowledge of the modular weights
$q^I_{sec}$ (see (3.2)).  They are equal to
1 for the untwisted matter fields, $\frac{2}{3}$
for the twisted {\bf (1,27,1)}-multiplets and 1 for the twisted
${\bf (1,1,\bar{3})}$-multiplets.  Furthermore, it can be
shown~[3] that $\ln \det \G_{moduli} = 6 K_M$.  Inserting this
information into (A.3) and
retaining only the dependence on the untwisted moduli yields
\beqa 2 \ln \det {\cal G}_{I \bar{J}} + 2 \sum_{sec} ( \prod_H
n_{sec}^H ) \ln \det \G_{sec} = 1632 K_M \eeqa 
Hence, the moduli dependent part of $N$ cancels out.
Note that, when supplementing (A.1) with the counterterms
introduced in the main part of this paper, the moduli
dependence of these counterterms cancels out as well.  Since
the introduction of these counterterms renders $N$
a scalar function of the untwisted moduli,
its moduli dependence cancels out in any coordinate system.
It is then tempting to conjecture that
the coefficient $c$ given in (A.2) represents the amount of
the $K$-terms
in (A.1), which is removed by the Green-Schwarz
mechanism in the following way.
The term proportional to $(W_{\alpha
\beta \gamma})^2$ in (5.13), when inserted into the Green-Schwarz
Lagrangian (5.18), contributes the following non-local term
$\Gamma_{GS}$ to the effective theory
\beq  \Gamma_{GS} = - \beta \tau \int d^4 \theta
W^{\alpha \beta \gamma} W_{\alpha \beta \gamma} \frac{1}{\Box}
D^2 {\kappa}^2 G + h.c.     \eeq 
By comparing (A.6) against (A.1) it follows
that the amount $c$ can be removed when choosing
\beq \tau = \frac{720}{\beta} \; \frac{1}{12} \;
\frac{1}{(4 \pi)^2} \eeq 
where $\beta$ is given in (5.23).

Next, we consider $\Gamma_{X^2}$ given in (4.21).  We rewrite
$\Gamma_{X^2}$ as follows
\beqa \Gamma_{X^2} &=& \frac{1}{96} \;
\frac{1}{(4\pi)^2}
\int d^4 \theta X^{\alpha} X_{\alpha}
\frac{1}{\Box} D^2 (p \kappa^2 K + U) + h.c. \eeqa  
where
\beq p = 744 \eeq 
and
\beqa U &=& \kappa^2 K ( - p - \sum_k ( \prod_H n_k^H )
- \sum_d ( \prod_H n_d^H ) - n_I - 1 + \sum_H n_V^H + 3 )  \nonumber\\
&+& 2 \ln \det {\cal G}_{I \bar{J}} + 2 \ln {\cal G}_S
+ 2 \sum_{sec} ( \prod_H n_{sec}^H ) \ln \det \G_{sec}  \eeqa 
In (A.10), we have included the contribution from the gravitino,
$+3$~[19].  The variation of $\Gamma_{X^2}$ under \Kahler\
transformations (4.25) was given in (4.27) and reads
\beqa             \delta_K \Gamma_{X^2} &=& \frac{1}{24}
\frac{1}{(4 \pi)^2} \int d^2
\theta \kappa^2 F(T,S) X^\alpha X_\alpha[
\sum_k (\prod_H n^H_k) + \sum_d (\prod_H n^H_d) + n_I \nonumber\\
&+& 1 - \sum_H n^H_V - 3 ] + h.c.  \eeqa 
In (A.11), we have included the contribution from the gravitino.
Inserting the information about the massless spectrum of $Z_3$
into (A.11) yields a non-vanishing result for the pure
\Kahler\ anomaly
\beq \delta_K \Gamma_{X^2} = \frac{1}{24} \frac{1}{(4 \pi)^2}
888 \int d^2 \theta \kappa^2 F(T,S) X^\alpha X_\alpha
+ h.c. \eeq 
In the standard (T,S)-coordinate system, it is readily seen
that the dependence of
$U$ on the untwisted moduli cancels out.  This can be seen
by comparing $U$, as given in (A.10), against $N$, as
given in (A.3).  Again, when supplementing $U$ with
the counterterms introduced in the main part of this
paper, the moduli dependence of these counterterms
cancels out as well.  Hence,
the moduli dependence of $U$ (supplemented by counterterms)
cancels out in any coordinate system.
It is then tempting to conjecture that
the coefficient $p$ in (A.9) represents the amount of the
$K$-terms in (A.8) which is removed by the Green-Schwarz
mechanism, as discussed in section 5, yielding
\beqa 3\sigma + 4\tau = - \frac{744}{\beta}\;
\frac{1}{2} \; \frac{1}{(4 \pi)^2}   \eeqa   

\section*{References}

\begin{enumerate}

\newcommand{\PR}[3]{Phys. Rev. {\bf #1} (19#2) #3}
\newcommand{\PL}[3]{Phys. Lett. {\bf #1} (19#2) #3}
\newcommand{\NuP}[3]{Nucl. Phys. {\bf #1} (19#2) #3}

\item P. Bin\'{e}truy, G. Girardi, R. Grimm and M. M\"{u}ller,
\PL{B189}{87}{83}; P. Bin\'{e}truy, G. Girardi and R. Grimm, LAPP
preprint TH-275/90.

\item E. Cremmer, B. Julia,
J. Scherk, S. Ferrara, L. Girardello and P. van Nieuwenhuizen,
\PL{B79}{78}{231}; \NuP{B147}{79}{105}; B. Zumino, \PL{B87}{79}
{203}; E. Cremmer, S. Ferrara,
L. Girardello and A. van Proyen, \PL{B116}{82}{231}; \NuP{B212}
{83}{413}; E. Witten and J. Bagger,
\PL{B115}{82}{202}; J. Bagger, \NuP{B211}{83}{302}.

\item G. L. Cardoso and B. A. Ovrut, University of Pennsylvania
preprint UPR-0481T (1991), to be published in Proc. Strings and
Symmetries 1991, Stony Brook, May 20-25, 1991; Proc. of
the International Europhysics Conference, Geneva, Switzerland,
July 1991.

\item L. J. Dixon, J. A. Harvey, C. Vafa and E. Witten,
\NuP{B261}{85}{678}; \NuP{B274}{86}{285}.

\item M. Cveti\v{c}, J. Louis
and B. A. Ovrut, \PL{B206}{88}{227}; M. Cveti\v{c}, J. Molera and
B. A. Ovrut, \PR{D40}{89}{1140}.

\item G. Moore and P. Nelson, Phys. Rev. Lett. {\bf 53} (1984) 1519;
J. Bagger, D. Nemeschansky and S. Yankielowicz,
\NuP{B262}{85}{478}.

\item L. J. Dixon, V. S. Kaplunovsky and J. Louis,
\NuP{B355}{91}{649}.

\item G. L. Cardoso and B. A. Ovrut, \NuP{B369}{92}{351}; J. P.
Derendinger, S. Ferrara, C. Kounnas and F. Zwirner, \NuP{B372}{92}
{145}.

\item J. Louis, PASCOS 1991 Proceedings, P. Nath ed.,
World Scientific 1991

\item I. Antoniadis, K. S. Narain and T. R. Taylor,
\PL{B267}{91}{37}.

\item J. Lauer, J. Mas and H. P. Nilles, \PL{B226}{89}{251};
R. Dijkgraaf, H. Verlinde and E. Verlinde, in Proc. Perspectives
in String Theory, Copenhagen 1987

\item M. B. Green and J. H. Schwarz, \PL{B149}{86}{117}.

\item M. M\"{u}ller, \NuP{B264}{86}{292}.

\item E. Witten, \PL{B155}{85}{151}.

\item S. Ferrara, D. L\"{u}st, A. Shapere and S. Theisen,
\PL{B225}{89}{363}; S. Ferrara, D. L\"{u}st and S. Theisen,
\PL{B233}{89}{147}; M. Cveti\v{c}, A. Font, L. E. Ib\'{a}\~{n}ez,
D. L\"{u}st and F. Quevedo, \NuP{B361}{91}{194}.

\item L. J. Dixon, V. S. Kaplunovsky and J. Louis,
\NuP{B329}{90}{27}.

\item S. L. Adler and W. Bardeen, \PR{182}{69}{1517}.

\item R. Garreis, M. Scholl and J. Wess, Z. Phys. {\bf C28}
(1985) 623.

\item L. Alvarez-Gaum\'{e} and P. Ginsparg, Annals of Phys. {\bf
161} (1985) 423.

\item S. Ferrara and M. Villasante, \PL{B186}{87}{85}; P.
Bin\'{e}truy, G. Girardi, R. Grimm and M. M\"{u}ller,
\PL{B195}{87}{389}; P. Bin\'{e}truy, G. Girardi and R. Grimm,
\PL{B265}{91}{111}.

\item G. Girardi and R. Grimm, \NuP{B292}{87}{181}.

\item L. E. Ib\'{a}nez and D. L\"{u}st, preprint CERN TH.6380/92;
I. Antoniadis, E.Gava and K.S. Narain, preprint CPTH-A162.0392,
IC/92/51(1992)
\end{enumerate}

\newpage

\section*{Figure Captions}

Figure 1: We present the anomalous three-point triangle graph for
matter fields $\phi^k$ running around the triangle loop.
\newline
Figure 2: We present the anomalous tree graph contribution from
the Chern-Simons and Green-Schwarz terms.

\end{document}